\documentclass[conference]{IEEEtran}

\usepackage{amsmath,amssymb,amsfonts}
\usepackage{algorithmic}
\usepackage{graphicx}
\usepackage{textcomp}
\usepackage{xcolor}
\usepackage{float}
\usepackage{subcaption}

\usepackage{color}

\usepackage[style=ieee, url=false, isbn=false, eprint=false, dashed=false]{biblatex} 
\def\BibTeX{{\rm B\kern-.05em{\sc i\kern-.025em b}\kern-.08em
    T\kern-.1667em\lower.7ex\hbox{E}\kern-.125emX}}

\begin{document}

\title{Synthesizing Compound Pulse Gadgets for Hamiltonian Simulation on Trapped-Ion Platforms}

\author{%
    \IEEEauthorblockN{
        Ria Patel$^{1}$,
        Masoud Hakimi Heris$^{2}$,
        Yuan Liu$^{1,2,3}$,
        Frank Mueller$^{1}$ 
}
\IEEEauthorblockA{$^{1}$Department of Computer Science,
North Carolina State University, Raleigh, NC 27695, USA}
\IEEEauthorblockA{$^{2}$Department of Electrical and Computer Engineering,
North Carolina State University, Raleigh, NC 27695, USA}
\IEEEauthorblockA{$^{3}$Department of Physics,
North Carolina State University, Raleigh, NC 27695, USA\\
Corresponding authors: Ria Patel (rpatel38@ncsu.edu),
Frank Mueller (fmuelle@ncsu.edu)}
}

\maketitle

\begin{abstract}
Standard gate-level transpilation introduces significant physical noise and overhead for high-precision quantum algorithms, such as the Quantum Singular Value Transformation (QSVT), on near-term trapped-ion hardware. Current compilers treat quantum operations as discrete units, forcing the physical control layer to execute highly fragmented laser pulses. To address this hardware-software disconnect, this work introduces a holistic pulse synthesis strategy that bypasses discrete gate-stitching to compile algorithms directly into continuous compound pulse gadgets. As a proof-of-concept, we target Hamiltonian simulation of the $H_2$ molecule, block-encoding the problem into a QSVT circuit to approximate the time-evolution operator $U = e^{-i H t}$ across 3 computational ions (2 system, 1 ancilla). We utilize the Gradient Ascent Pulse Engineering (GRAPE) algorithm to generate these compound gadgets and evaluate our methodology using noisy Lindblad master equation simulations. Preliminary observations indicate that the proposed strategy achieves significant temporal compression, reducing the total pulse schedule duration compared to standard compilers. Furthermore, synthesizing operations holistically eliminates the control-layer latency associated with discrete pulse lookup overhead. By streamlining the physical control schedule, this methodology offers a promising pathway to execute operations faster, highlighting the potential for compound gadgets to increase the computational depth achievable within fundamental $T_2$ decoherence limits.

\end{abstract}


\vspace{-3mm}
\section{Introduction}

Standard gate-level transpilation introduces significant overhead,
severely bottlenecking high-precision algorithms like the Quantum
Singular Value Transformation (QSVT)
\cite{gilyenQuantumSingularValue2019a,
  martynGrandUnificationQuantum2021}. QSVT provides a unified
mathematical paradigm for executing optimal quantum algorithms by
interleaving a block-encoded non-unitary matrix $U_A$ with
parameterized projector rotations $R_\phi$ to apply polynomial
transformations to the embedded singular values
\cite{lowOptimalHamiltonianSimulation2017, joven2026,
  toyoizumiHamiltonianSimulationUsing2024}. While mathematically
elegant and theoretically optimal
\cite{motlaghGeneralizedQuantumSignal2024}, QSVT produces
exceptionally deep circuits. Current compiler stacks
\cite{bqskit_doecode_58510} treat these dense, alternating blocks as
discrete mathematical abstractions, resulting in massive circuit depth
that is at odds with the short decoherence limits of near-term hardware.

Mapping these discrete operations to trapped-ion hardware forces the
physical control layer to execute sharp, highly fragmented laser pulses. 
Trapped-ion systems encode discrete-variable (DV) logic in atomic energy levels while utilizing continuous-variable (CV) motional modes (phonons) for multi-qubit 
entangling operations like Mølmer-Sørensen (MS) gates. This exposes 
computations to $T_2$ dephasing and anomalous motional heating. Standard 
discrete transpilation exacerbates heating by executing highly discontinuous 
square waves that require massive peak laser power and sub-microsecond optical 
switching. These sharp time-domain edges leak broadband frequency noise 
to outside the computational subspace, rapidly degrading the quantum state 
\cite{kang2021batch}.

While advanced synthesis frameworks like BQSKit \cite{bqskit_doecode_58510} 
excel at algebraic unitary resynthesis to minimize discrete gate counts, 
they ultimately rely on executing fragmented schedules of isolated operations. 
Much of current optimal control research also remains focused on optimizing 
these individual gate primitives in isolation \cite{dalviGraphBasedPulseRepresentation2024,
khanejaOptimalControlCoupled2005c}. This leaves a significant gap between 
algorithmic theory and physical execution.


This work bridges that gap. We introduce a holistic compilation methodology 
targeting the physical control layer directly. Bypassing discrete gate 
representations, we utilize numerical optimal control (GRAPE) to synthesize 
entire multi-qubit QSVT algorithmic blocks into continuous, hardware-native 
pulse gadgets. As a proof-of-concept, we target a 3-ion Hamiltonian simulation 
of the $H_2$ molecule \cite{PySCF, qiskit2024}. By synthesizing these operations 
holistically, we strictly minimize the pulse schedule duration and investigate 
the viability of executing deep algorithms without triggering catastrophic 
motional heating.

In summary, we introduce a pulse-level compilation strategy, supported
by the following core contributions:
\begin{itemize}
\item \textbf{Compound Pulse Gadgets:} We propose a generalized
  compilation methodology that bypasses discrete gate abstractions,
  synthesizing high-level algorithmic specifications directly into
  physical control schedules. We demonstrate this technique by
  targeting Hamiltonian simulation of the $H_2$ molecule via QSVT,
  pairing interleaved unitaries into continuous compound pulse
  gadgets.
\item \textbf{Temporal Compression:} We demonstrate that our
  gadget-based GRAPE optimization strategy significantly reduces the
  total algorithmic execution duration. By compiling directly to
  continuous physical waveforms, we eliminate the temporal bloat
  inherent to discrete modular gate-stitching. This temporal
  compression highlights a promising pathway to increase the
  computational depth achievable within fundamental hardware
  decoherence limits.
\item \textbf{Elimination of Lookup Overhead:} We show that
  synthesizing a 
  few
  high-level operations (unitaries) holistically removes
  the need to issue a large number of sequential gate-primitive memory
  lookups during schedule generation. By bypassing this fragmented
  assembly process, our methodology drastically reduces control-layer
  latency, providing a highly hardware-efficient execution paradigm
  evaluated through noisy environmental simulations.
\end{itemize}

\begin{figure*}[htbp] 
     \centering
     \includegraphics[width=0.70\textwidth]{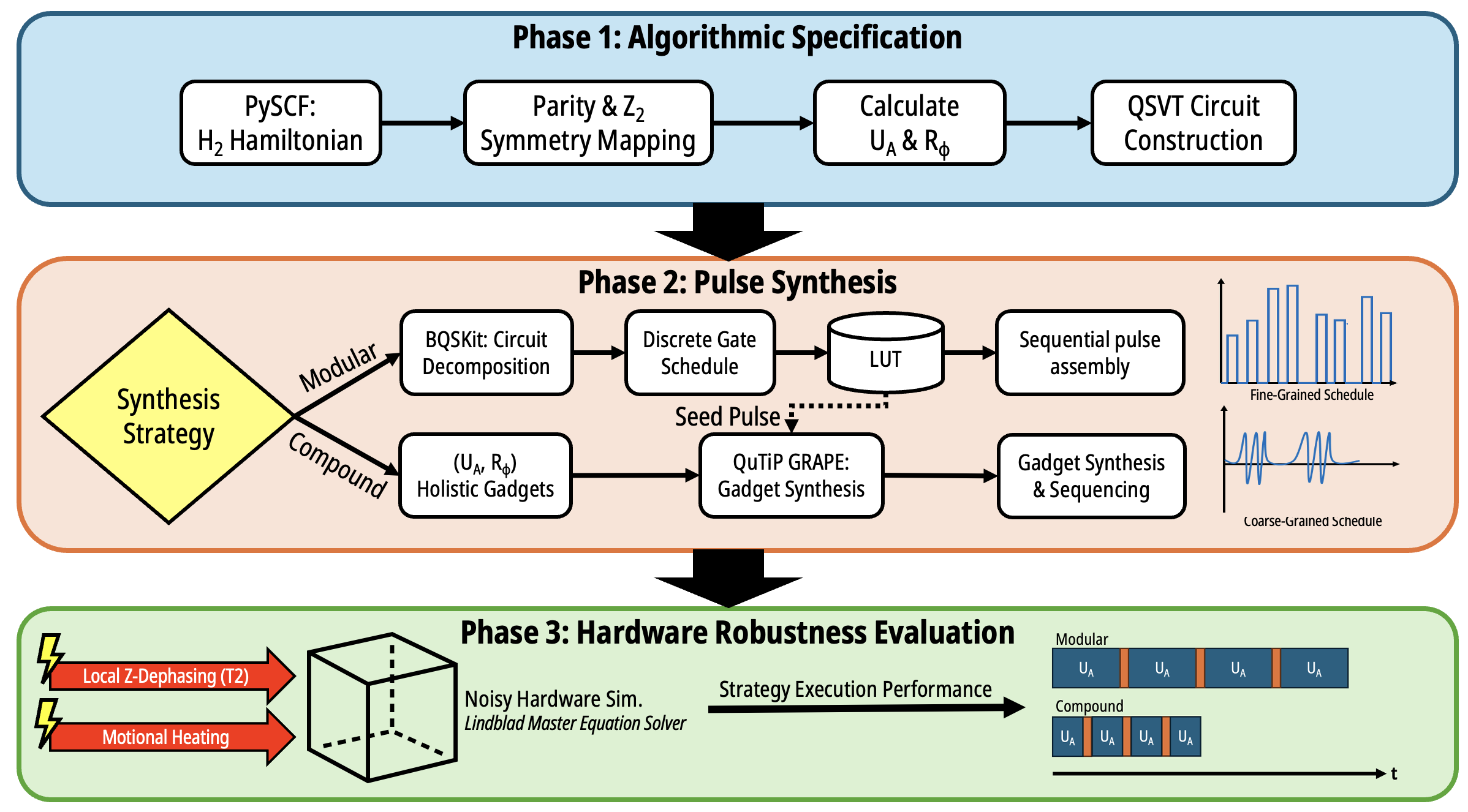}
     \caption{The pulse-level compilation pipeline. Phase 1 maps the
       $H_2$ Hamiltonian to a 3-ion QSVT specification. Phase 2
       contrasts standard modular transpilation (sequential pulse
       assembly via LUT memory lookups) against the proposed holistic
       gadget synthesis. Phase 3 evaluates strategy execution
       performance, benchmarking the temporal compression of the
       generated schedules against static hardware noise using
       open-system Lindblad simulations.
     }
     \vspace{-1.5em}
     \label{fig:flowchart}
\end{figure*}

\section{Methodology}

The objective of this work is to evaluate the efficiency and viability
of our compound pulse gadget synthesis methodology on trapped-ion
platforms. We evaluate this approach using a QSVT circuit that
performs Hamiltonian simulation of the $H_2$ molecule. Our strategy is
designed to bridge the gap between high-level algorithmic
specifications and low-level physical control schedules. Our
methodology (see Figure \ref{fig:flowchart}) is split into three main phases: algorithmic
specification, pulse synthesis, and hardware robustness evaluation.

\subsection{Algorithmic Specification and Strategy Design}
The proposed methodology serves as the central orchestration layer for
translating high-level quantum algorithms into physical control
signals. We target Hamiltonian simulation to approximate the
time-evolution operator $U = e^{-i H t}$. To construct the target
problem, our approach utilizes the PySCF chemistry driver \cite{PySCF}
alongside Qiskit \cite{qiskit2024} to compute the Hamiltonian for the
$H_2$ molecule (STO-3G basis), which is then mapped from 4
spin-orbitals to 2 logical qubits via tapering by symmetry
exploitation.

We block-encode this normalized 2-qubit Hamiltonian into a larger
unitary $U_A$, which requires one additional ancilla qubit for signal
processing. Consequently, the final algorithm is mapped to exactly 3
computational trapped ions. Using a specified target error bound,
$\epsilon$, the methodology then calculates the necessary QSVT phase
angles $\phi$.

For the $H_2$ Hamiltonian, the resulting QSVT polynomial is of degree
4. The logical circuit manifests as a sequence of interleaved
block-encodings $U_A$ and projector-controlled phase shifts $R_\phi$.

To transition this mathematical operator sequence into a physical
instruction stream, it is crucial to define the target computational
subspace. Trapped-ion hardware consists of both discrete-variable (DV)
internal atomic states and continuous-variable (CV) collective
motional modes. While our strategy strictly compiles DV quantum logic
(qubits), it avoids executing this logic via standard discrete-time
gates. Instead, the technique deploys continuous physical control
waveforms that actively suppress unwanted excitations in the CV
motional bath. To evaluate this approach, we coordinate the
compilation through two distinct synthesis strategies: Modular and
Compound.

\begin{figure*}[htbp] 
     \centering
     \includegraphics[width=0.75\textwidth]{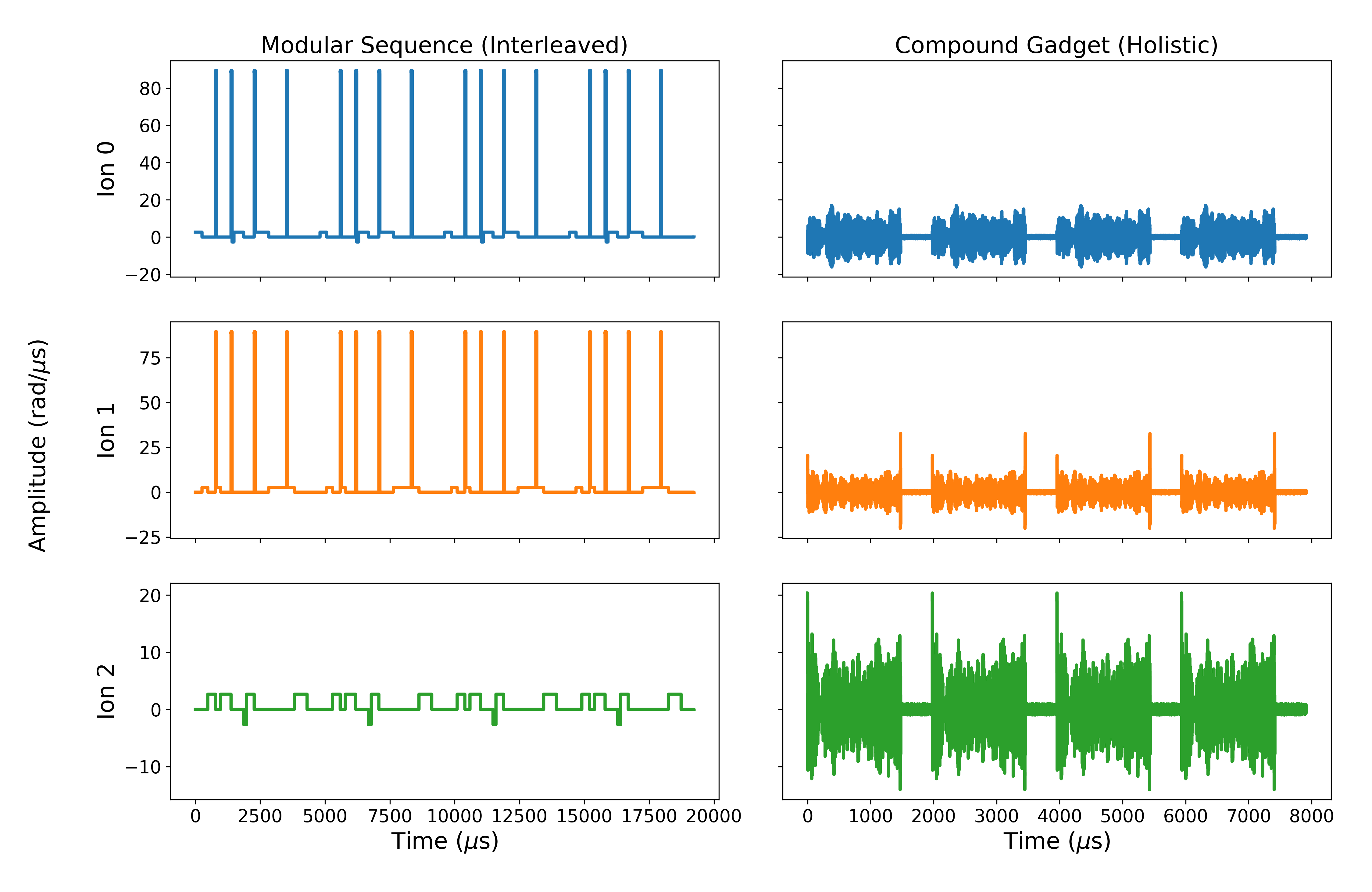}
     \caption{Comparison of physical control schedules constructed by
       the modular baseline and the proposed compound gadget synthesis
       strategy for QSVT-based Hamiltonian simulation.}
     \vspace{-4mm}
     \label{fig:sequence_comparison}
\end{figure*}

\subsection{Pulse Synthesis Strategies}
Before detailing the specific compilation methods, it is critical to
establish the physical boundaries of the synthesis environment. Both
the modular baseline and the proposed compound strategy generate pulse
schedules under a noiseless assumption. The optimizers focus strictly
on achieving the target discrete-variable (DV) unitary logic in an
ideal vacuum. Environmental noise and ambient continuous-variable (CV)
heating are purposely excluded from the synthesis phase to ensure the
matrix exponentials remain computationally tractable.

\subsubsection{Modular Synthesis Strategy (Baseline)}
To establish a baseline, we simulate a standard transpilation
approach. The target unitaries are decomposed into a discrete schedule
of 144 native trapped-ion operations (RXX entangling gates and RZ/RX
single-qubit rotations) using the BQSKit compiler
\cite{bqskit_doecode_58510}.

To translate this logical schedule into physical control signals, we
utilize a pre-calculated Mølmer-Sørensen (MS) gate primitive. This
physical RXX waveform is generated using pulse optimization software
provided by the Duke Quantum Center \cite{duke_ms_gate_repo}. The gate
primitive is synthesized based on empirical trapped-ion hardware
parameters and robust modulation schemes \cite{kang2021batch}, derived
under an ideal noiseless assumption. This optimized MS waveform, along
with generated single-qubit RX control waves, is stored in a Look-Up
Table (LUT). Notably, RZ rotations are compiled as virtual gates,
executed instantaneously via software phase tracking rather than
physical laser pulses. The full physical control sequence is then
constructed by retrieving the required waveforms from the LUT and
stitching them sequentially according to the BQSKit instruction
schedule. Even with 0-duration virtual Z-rotations, this strategy
treats every physical operator in isolation. For a single QSVT block,
this results in highly fragmented control signals and significant
execution bloat as the classical controller processes 144 sequential
memory lookups.

\subsubsection{Compound Gadget Synthesis Strategy}
This strategy represents the core contribution of this work. Instead
of treating operators as discrete units, our methodology pairs the
interleaved $(U_A, R_\phi)$ blocks into holistic compound pulse
gadgets.

To synthesize the physical control waveforms for these gadgets, the
pipeline utilizes the Gradient Ascent Pulse Engineering (GRAPE)
algorithm via the QuTiP software \cite{qutip5,
  Li2022pulselevelnoisy}. Operating exclusively within the noiseless
physical model, the L-BFGS-B numerical optimizer searches for a
continuous drive waveform that enacts the target DV
unitary. Crucially, the optimizer generates these schedules
dynamically, bypassing the physical limitations and fragmented
assembly process of the modular baseline. The synthesized compound
gadgets are then sequenced to form the complete, temporally compressed
physical execution track for the QSVT algorithm.

\subsection{Preliminary Robustness Evaluation}
Pulse synthesis is performed in a frictionless, noiseless
environment. To verify that the synthesized pulse schedules are
physically viable, we conduct a preliminary experimental evaluation
using a noisy Lindblad master equation \cite{qutip5, Li2022pulselevelnoisy}
. This
experimental setup is designed to stress-test the generated control
schedules against realistic environmental noise without requiring
immediate physical hardware access.

This methodology models a trapped-ion chain consisting of 3
computational ions coupled to a single shared motional mode. To
maintain computational tractability while providing sufficient
mathematical headroom, the bosonic Fock space of the shared
continuous-variable (CV) mode is appropriately truncated. To capture
realistic environmental noise, we inject local Z-dephasing ($T_2$
noise) and motional anomalous heating into the noisy simulation.

To evaluate the systems-level advantages of our strategy, both
compilation approaches are tasked with executing the identical logical
QSVT block. Rather than constraining the simulations to an arbitrary
physical time window, we allow the discrete baseline and the compound
gadget strategy to run for the full duration required by their
respective physical control schedules. This experimental setup allows
us to directly measure the temporal compression achieved by holistic
synthesis, and to evaluate how shrinking the physical execution window
naturally improves state preservation against the static $T_2$ noise
floor.

\section{Results and Discussion}
While these performance gains are explicitly demonstrated for a 3-ion Hamiltonian simulation of the $H_2$ molecule, we hypothesize that the underlying benefits observed (temporal compression and reduced lookup latency) will scale to more complex Hamiltonians. Validating this work against larger molecules like $LiH$ is an objective for future work.

\textbf{Observation 1: Compound pulse gadgets achieve significant
  temporal compression.}  As illustrated in Figure
\ref{fig:sequence_comparison}, the standard modular baseline is
severely bottlenecked by the sequential execution of isolated gate
primitives. By synthesizing the QSVT block holistically, the compound
gadget strategy significantly compresses the total execution
window. Because our methodology compresses the same logical operations
into a much shorter physical duration, it drastically increases the
computational depth achievable within the fundamental $T_2$
decoherence ceiling. Ultimately, this temporal compression allows the
hardware to execute significantly more algorithmic blocks in the time
than it would take the modular baseline to complete a single
fragmented schedule.

\textbf{Observation 2: Holistic synthesis eliminates discrete pulse
  lookup overhead.}  Rather than executing a highly fragmented
schedule of discrete gates, our methodology elevates the compilation
boundary to the algorithmic block level. We synthesize a holistic,
temporally compressed pulse gadget for the $(U_A,R_\phi)$ unit. The
classical controller then executes a coarse-grained schedule,
sequencing these gadgets at the algorithmic boundaries. This reduces
the classical instruction overhead from hundreds of discrete gate
fetches down to just $d$ macro-block fetches (where $d$ is the QSVT
polynomial degree), vastly minimizing control-layer latency while
preserving the temporal compression of the internal block.

Ultimately, these preliminary findings suggest that pushing
compilation directly to the pulse-level provides a highly
hardware-efficient methodology. By maximizing temporal compression and
eliminating lookup latency, this compound pulse synthesis strategy
makes the schedules robust to baseline hardware noise, establishing a
clear pathway to increase computational depth and execute larger
algorithmic blocks on current trapped-ion systems.

\begin{figure}[htbp] 
     \centering
     \includegraphics[width=\linewidth]{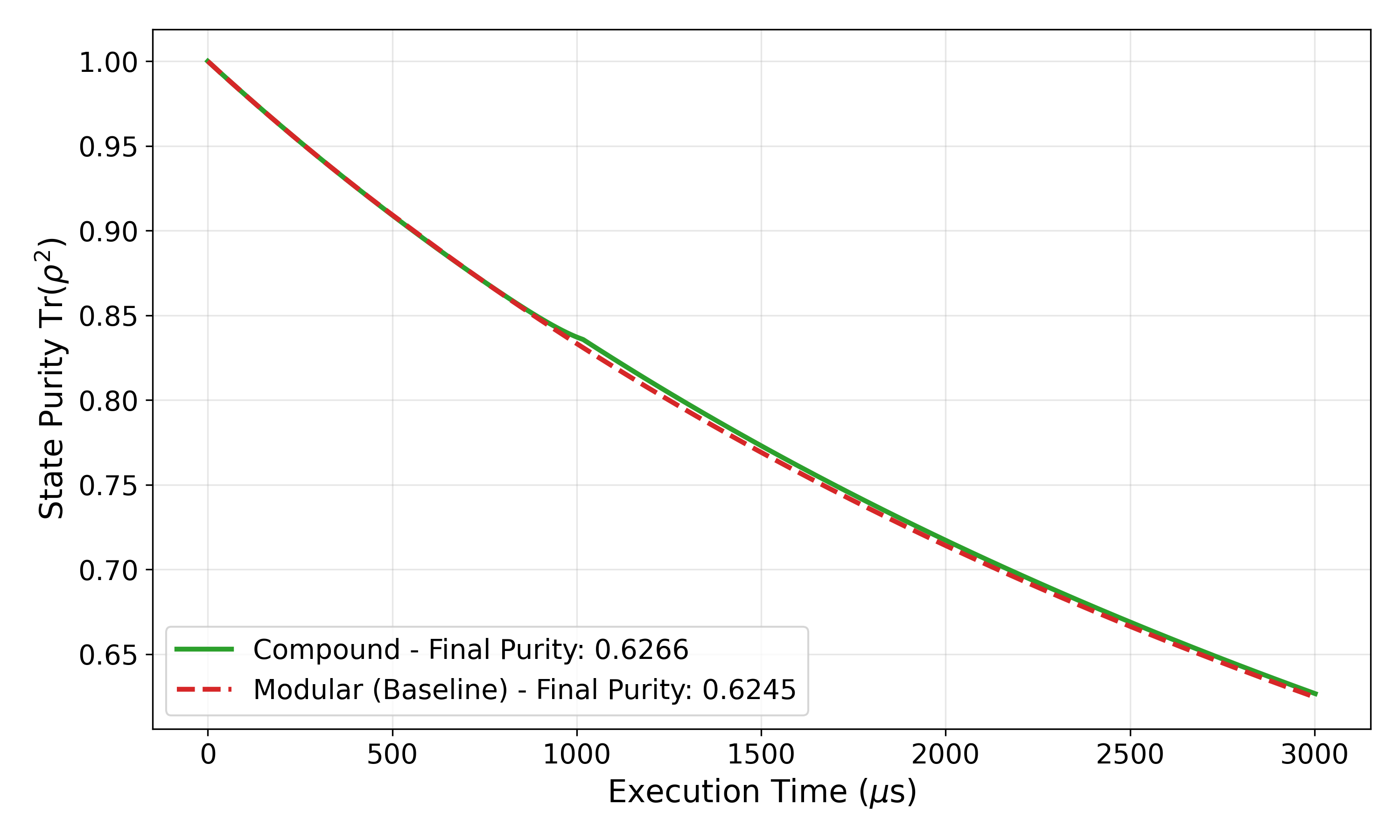}
     \caption{Hardware resilience evaluated via open-system purity decay ($Tr(\rho^2)$) over a 3 ms execution window, comparing the baseline modular stitching against the proposed compound compilation.}
     \vspace{-4mm}
     \label{fig:purity_comparison}
\end{figure}

\section{Conclusion and Future Work}

In this work, we introduced a pulse-level compilation methodology that
bypasses standard discrete gate-stitching and synthesizes QSVT
algorithms directly into continuous compound pulse gadgets. By
targeting Hamiltonian simulation for the $H_2$ molecule on trapped-ion
hardware, our preliminary experiments demonstrate that holistic pulse
synthesis achieves significant temporal compression compared to a
discrete modular baseline. Furthermore, this strategy drastically
streamlines the physical instruction schedule, eliminating the
control-layer latency associated with sequential memory lookups. By
discarding the rigid abstractions of discrete gates, our methodology
generates highly efficient physical control schedules that maximize
computational depth before the system reaches the static hardware
noise floor.

Building upon this proof-of-concept, future research will actively
scale this compilation strategy to process significantly deeper
algorithmic circuits. We plan to deploy the GRAPE optimization
pipeline on an institutional computing cluster to utilize
computational accelerator resources for synthesizing larger continuous
compound gadgets. 
To address the exponential scaling bottleneck of holistic synthesis for wider systems, we also intend to explore unitary partitioning and cutting paradigms to subdivide massive algorithmic blocks into computationally tractable optimal control targets.
We also plan to investigate time-optimal control
boundaries within the numerical optimizer to push temporal compression
to its theoretical limits and maximize the number of algorithmic
blocks that can be executed within strict $T_2$ decoherence
windows. Additionally, we aim to augment the compiler's cost function
with constraints to ensure the generated schedules remain robust
against complex hardware-specific noise channels without relying on
discrete error suppression. Finally, we will transition from simulated
robustness evaluations to empirical hardware validation, testing the
systems-level advantages of our compiler directly on physical
trapped-ion processors.

\section*{Acknowledgments}
This work was supported in part by OSI-2531350,
OMA-2120757 and PHY-2325080.
Yuan Liu and Frank Mueller were also supported in part by the
U.S. Department of Energy, Office of Science, Advanced Scientific
Computing Research, under contract number DE-SC0025384.

\printbibliography

\end{document}